\documentclass[12pt]{iopart}

\usepackage{iopams}  
\usepackage{graphicx}
\usepackage{citesort}
\usepackage{bm}
\usepackage{times}

\graphicspath{{.}{./EPS/}}

\newcommand*{\vek}[1]{{\ensuremath{\bm{\mathrm{#1}}}}}
\newcommand*{\omegac}[1][]{\omega_{{\mathrm c}#1}}
\newcommand* {\expect}[1]{\ensuremath{\left\langle {#1} \right\rangle}}

\begin{document}

\title[Magnetic focusing of charge carriers from spin-split bands]{Magnetic focusing of
charge carriers from spin-split bands: Semiclassics of a \textit{Zitterbewegung\/} effect}

\author{U Z\"ulicke$^1$, J Bolte$^2$ and R Winkler$^3$}

\address{$^1$~Institute of Fundamental Sciences and MacDiarmid Institute for
Advanced Materials and Nanotechnology, Massey University, Palmerston North,
New Zealand}
\address{$^2$~Institut f\"ur Theoretische Physik, Universit\"at Ulm,
Albert-Einstein-Allee~11, D-89069 Ulm, Germany}
\address{$^3$~Department of Physics, Northern Illinois University, DeKalb,
IL 60115, USA}
\ead{u.zuelicke@massey.ac.nz}

\date{\today}

\begin{abstract}
We present a theoretical study of the interplay between cyclotron motion and spin splitting
of charge carriers in solids. While many of our results apply more generally, we focus especially
on the Rashba model describing electrons in the conduction band of asymmetric
semiconductor heterostructures. Appropriate semiclassical limits are distinguished that describe
various situations of experimental interest. Our analytical fomulae, which take full account of
Zeeman splitting, are used to analyse recent magnetic-focusing data. Surprisingly, it turns out
that the Rashba effect can dominate the splitting of cyclotron orbits even when the Rashba and
Zeeman spin-splitting energies are of the same order. We also find that the origin of
spin-dependent cyclotron motion can be traced back to \textit{Zitterbewegung\/}-like oscillatory
dynamics of charge carriers from spin-split bands. The relation between the two phenomena is
discussed, and we estimate the effect of \textit{Zitterbewegung\/}-related corrections to
the charge carriers' canonical position.
\end{abstract}

\submitto{\NJP}
\maketitle

\section{Introduction}

\begin{figure}[b]
\begin{indented}
\item[] \includegraphics[width=8cm]{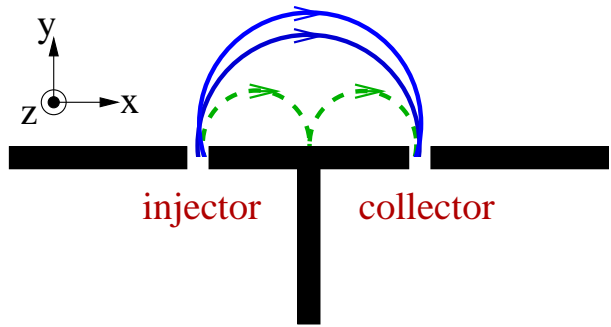}
\caption{\label{fig:FocSetup}Schematic setup for a magnetic-focusing experiment. Current is
passed through the injector contact, and the voltage in the collector contact is monitored as a
function of a magnetic field applied in the $z$ direction. The latter forces charge carriers to
move on cyclotron orbits in the $xy$ plane. Particles injected on trajectories starting out
sufficiently close to parallel to the $y$ direction will be focused into the collector contact when
the contact separation equals an integer multiple of their cyclotron-orbit diameter. As a result,
peaks are observed in the collector voltage at the corresponding magnetic-field values.}
\end{indented}
\end{figure}

Magnetic focusing of ballistic charge carriers in metals~\cite{sharvMagFocus,tsoiMagFocus}
and semiconductors~\cite{magFocus,magFocRev,here:apl:92} has been successfully used to
elucidate fundamental materials properties such as the shape of the Fermi
surface~\cite{FSmagFoc,here:surf:94}, Andreev reflection in superconductor/normal-metal hybrid
structures~\cite{focus1,focus3}, surface crystallography~\cite{surfMagFoc}, phase coherent
transport~\cite{magFocus}, and emergent quasiparticles in the fractional-quantum-Hall
regime~\cite{CFmagFoc}. The fundamental setup of a magnetic-focusing experiment is quite
simple; see \Fref{fig:FocSetup}.
It requires sufficiently ballistic transport between two fixed, co-linear (injector and collector)
contacts. A large number of injected-particle trajectories converge at the collector every time
the contact separation $L$ equals an integer multiple of the cyclotron diameter
$2 r_{\mathrm c}$. Peaks occurring in the measured collector voltage at concomitant magnetic
fields are the experimental signature for magnetic focusing. Recently, magnetic-focusing
trajectories of electrons in a two-dimensional semiconductor heterostructure have been imaged
directly using scanning-probe microscopy~\cite{west:07}.

Several experiments~\cite{spinPolMagFoc,rokh:prl:04,rokh:prl:06,murph:physe:06,here:aipproc:07}
have investigated the possibility to spatially separate charge carriers belonging to spin-split
bands using the magnetic-focusing technique, which may have ramifications for the emerging
field of spin electronics~\cite{sciencerev,zut:rmp:04}. The desire to use
magnetic-field-independent, spin-orbit-induced spin splitting for this
purpose~\cite{rokh:prl:04,rokh:prl:06,murph:physe:06,here:aipproc:07} is fuelling renewed
interest~\cite{usaj:prb-rc:04,usaj:prb:04,valin:prb:06,usaj:prb:07,usaj:07} in the theoretical study
of spin-orbit effects in the semiclassical
regime~\cite{litt:pra:91,litt:pra:92,frisk:ann:93,shay:prl:98,shay:prb:99,roess:epl:03,bolte:prl:98,bolte:ann:99,bolte:ahp:05,brack:jpa:02,plet:prl:02,plet:jpa:03,kepp:prl:02}.
We present a critical analysis of the interplay between spin-orbit coupling and cyclotron motion
in both quantum and semiclassical regimes, with particular emphasis on spin-dependent
magnetic focusing. We also discuss this effect in the context of another topic of great current
interest, namely {\em Zitterbewegung\/} in solid-state
systems~\cite{zawa:prb:05,schlie:prl:05,schlie:prb:06,jozsef:prb:06,zawa:prb:06,uz:prb:07,zawa:jpcm:07},
and comment on the (un-)suitability of proposed intuitive interpretations in terms of a
spin-dependent focusing field~\cite{rokh:prl:04} or a spin-dependent Lorentz
force~\cite{nikolic:prb:05,shen:prl:05,shen:prb:06}.

This article is organised as follows. We start by reviewing semiclassical theories that have been
developed for systems with finite spin splitting. These theoretical approaches are then applied to
describe cyclotron motion of charge carriers in two-dimensional (2D) heterostructures subject to
both Zeeman and Rashba~\cite{byra:jetplett:84,byra:jpc:84} spin splittings. Our results are used
to analyse recent experimental data obtained from p-type GaAs~\cite{rokh:prl:04} and n-type
InSb~\cite{murph:physe:06,here:aipproc:07} samples. Subsequently, we discuss the intricate
connection between spin-dependent magnetic focusing and \textit{Zitterbewegung\/} of charge
carriers from spin-split bands~\cite{schlie:prl:05,schlie:prb:06,jozsef:prb:06,uz:prb:07}. Conclusions
are given in the final Section.

\section{Basic aspects of semiclassics in presence of spin splitting}

Non-relativistic single-particle Hamiltonians with spin splitting can generally be written
in the form~\cite{bolte:ann:99,bolte:ahp:05,brack:jpa:02,plet:prl:02,plet:jpa:03}
\begin{equation}\label{eq:splitHam}
{\mathcal H} = {\mathcal H}_0(\vek{r}, \vek{p}) +
\vek{\mathcal B} (\vek{r}, \vek{p}) \cdot \vek{S} \quad .
\end{equation}
Here we denote the particle's position, momentum, and spin angular-momentum
operators by $\vek{r}$, $\vek{p}$, and $\vek{S}=\hbar\vek{s}$,
respectively,
where $\vek{s}$ is the vector of generators for SU(2) rotations in the spin-$s$ representation,
and $\hbar$ is the Planck constant. The vector operator $\vek{\mathcal B}$ represents
an effective magnetic field that contains, in general, contributions due to the Zeeman
effect and spin-orbit coupling. Note that $\vek{\mathcal B}$ has the
dimensionality 1/time. In the following, we distinguish three possible semiclassical limits.
To keep our notation uncluttered, we do not explicitly distinguish quantum-mechanical
operators from their associated semiclassical phase-space symbols. We will clearly
separate the discussion of purely quantum and semiclassical properties to avoid
any possible confusion arising from this simplification.

\subsection{Precessing-spin semiclassics}

Representing a truly quantum-mechanical correction, the second (spin-splitting) term
in \eref{eq:splitHam} vanishes in the usual semiclassical limit that is defined
as $\hbar \to 0$ such that $|\vek{s}| = |\vek{S}|/\hbar$
remains constant. As a result, the semiclassical orbital dynamics is unaffected by the
spin degree of freedom~\cite{bolte:prl:98,bolte:ann:99,bolte:ahp:05}. The classical
trajectory $\{\vek{r}(t),\vek{p}(t)\}$, as determined by ${\mathcal H}_0$, prescribes the
dynamics of the classical spin $\vek{s}$ via a precession-type equation of motion
\begin{equation} \label{eq:precWeak}
\dot\vek{s} = \vek{\mathcal B}\left( \vek{r}(t), \vek{p}(t) \right) \times \vek{s}
\quad ,
\end{equation}
i.e., we can interpret $|\vek{\mathcal B}|$ as the precession frequency.
This type of semiclassics has also been called a {\em weak-coupling
limit\/}~\cite{bolte:ann:99,bolte:ahp:05,brack:jpa:02,plet:prl:02,plet:jpa:03}. Physically, it
corresponds to the regime of a perfectly classical, spin-independent orbital motion with
an associated trajectory-dependent precession of a classical
spin~\cite{bolte:prl:98,bolte:ann:99,bolte:ahp:05,brack:jpa:02}.

In many experimentally relevant situations the appropriate semiclassical description will be 
of the precessional type discussed here. This has, e.g., been shown for 
anomalous magneto-oscillations~\cite{kepp:prl:02}. However, there are experimentally accessible 
regimes where the spin dynamics actually influences the orbital motion. Spin-dependent cyclotron
motion represents a pertinent example~\cite{rokh:prl:04,murph:physe:06,here:aipproc:07}. The
semiclassical description of such situations is desirable, motivating the consideration of alternative
schemes for performing the classical limit. We proceed to discuss two of these,
both of which correspond to a strong-coupling-type semiclassics.

\subsection{Spin-orbit-intertwined semiclassics}

One possibility to keep the spin-splitting term in the Hamiltonian~(\ref{eq:splitHam}) 
finite in the semiclassical limit $\hbar \to 0$ is to simultaneously require $|\vek{S}|$ to
remain constant, which implies $|\vek{s}| \to \infty$. The time evolution of the spin state
will then affect orbital dynamics and \textit{vice versa\/}. The resulting set of
semiclassical equations of motion is given by~\cite{bolte:ahp:05} 
\numparts
\begin{eqnarray}\label{eq:strong}
\dot\vek{r}  &=& \vek{\nabla}_{\vek{p}} {\mathcal H} \quad , \\
\dot\vek{p} &=& - \vek{\nabla}_{\vek{r}} {\mathcal H} \quad , \\
\dot\vek{S} &=& \vek{\mathcal B}( \vek{r}, \vek{p}) \times \vek{S}
\quad .
\end{eqnarray}
\endnumparts
Hence, in this case, the spin and orbital dynamics are mutually affecting each other.
These equations of motion have previously occured in~\cite{plet:prl:02,plet:jpa:03},
but without noticing that they only provide the leading semiclassical dynamics
when $|\vek{s}| \to \infty$ is considered in addition to $\hbar \to 0$. Moreover, 
in~\cite{plet:prl:02,plet:jpa:03} the notion of an extended phase space was introduced,
intending to stress the role of spin as an independent classical dynamical variable.
Kinematically, the same phase space arises in the precessing-spin semiclassics; however,
in that context the spin-orbit dynamics are not of a Hamiltonian form because the spin
dynamics is driven by the orbital motion, without any
feedback of the spin dynamics on the orbital motion.

\subsection{Adiabatic-spin semiclassics}

An alternative, in some sense
very-strong-coupling~\cite{litt:pra:91,litt:pra:92,frisk:ann:93,bolte:ann:99,brack:jpa:02,plet:jpa:03}
semiclassics is obtained by letting $\hbar\to 0$ while keeping the spin \emph{projection}
frozen to a quantised value $S_z = \hbar s_z$ with respect to the instantaneous direction
of $\vek{\mathcal B}$. Then the associated orbital dynamics is governed by the Hamiltonian
\begin{equation} \label{eq:adiabat}
{\mathcal H}_{s_z} = {\mathcal H}_0(\vek{r}, \vek{p})
+ S_z \, |\vek{\mathcal B}(\vek{r}, \vek{p})| \quad .
\end{equation}
For each possible value $S_z$, ranging over $S, S -\hbar, \dots, -S$, a generally
different classical trajectory is obtained. In contrast to the two cases discussed above,
here the spin degree of freedom itself has no dynamics; it just introduces a Berry-phase-like
contribution to the orbital motion~\cite{yuli:prl:93}. To avoid the problem of mode-conversion,
this approach is restricted to the case
$|\vek{\mathcal B} (\vek{r}, \vek{p})| > 0$ for every point along a trajectory~\cite{litt:pra:91,litt:pra:92}.

\section{Application to the Landau-Rashba model}

To be specific, we consider the Landau-Rashba model~\cite{byra:jpc:84,byra:jetplett:84} that
describes spin $s=1/2$ conduction-band electrons in an asymmetric 2D heterostructure subject to
a perpendicular magnetic field $\vek{B}=\vek{\nabla}_{\vek{r}} \times \vek{A} \equiv B\,
\vek{\hat z}$. It is of the form given in \eref{eq:splitHam}, with
\numparts
\begin{eqnarray}
{\mathcal H}_0(\vek{r}, \vek{p}) &=& \frac{1}{2 m} \left[ \vek{p} + e \vek{A}(\vek{r})
\right]^2 \quad , \\  \label{eq:laraB}
\vek{\mathcal B}(\vek{r}, \vek{p}) &=& \frac{g}{2} \, \omegac[0] \, \vek{\hat z} + \alpha
\left[ \vek{p} + e \vek{A}(\vek{r}) \right] \times \vek{\hat z} \quad .
\end{eqnarray}
\endnumparts
Here $m$ and $g$ are the effective mass and Land\'{e} factor of the
2D electrons, $\omegac[0] \equiv eB/m_0$ with $m_0$ the electron mass in vacuum,
and $\alpha$ characterises the strength of the Rashba spin
splitting~\cite{byra:jpc:84,byra:jetplett:84}. The 2D heterostructure growth direction is
taken as the Cartesian $z$ axis; with $\vek{\hat z}$ being the associated unit vector.

\subsection{Quantum solution: Jaynes-Cummings model}\label{JCmodelSec}

A complete quantum solution of the Landau-Rashba model is available~\cite{rashba},
as it is equivalent~\cite{plet:jpa:03} to the exactly soluble Jaynes-Cummings (JC)
model~\cite{jcmodel} in the rotating-wave approximation. (The JC model 
describes coupling of a harmonic oscillator, here associated with the Landau levels,
to a two-level system, here represented by the electronic spin degree of freedom.)
Using techniques developed in theoretical quantum optics~\cite{theoQOpt}, we
recently obtained~\cite{uz:prb:07} the exact Heisenberg time evolution of spin and
position operators. The perpendicular-to-the-plane spin component can be separated
into two parts, $S_z(t)={\bar S}_z + {\tilde S}_z (t)$, with time-independent and
oscillating parts given by
\numparts
\begin{eqnarray} \label{eq:barS}
{\bar S}_z &=& \frac{\hbar^2}{4} \frac{\mathcal{B}^{\mathrm{JC}}_z}
{\vek{\mathcal B}^{\mathrm{JC}} \cdot \vek{S}}
= \frac{\hbar}{2} \left( \frac{g}{2} - \frac{m_0}{m} \right)
\frac{\hbar\omegac[0]} {2\vek{\mathcal B}^{\mathrm{JC}} \cdot \vek{S}} \quad , \\[1ex] \label{eq:tildeS}
{\tilde S}_z (t) &=&  \left( S_z - {\bar S}_z \right) \, \exp \left( - 2 i t \,
\vek{\mathcal B}^{\mathrm{JC}} \cdot \vek{S}/\hbar \right)
\quad .
\end{eqnarray}
\endnumparts
Here $\vek{\mathcal B}^{\mathrm{JC}} = \vek{\mathcal B}(\vek{r}, \vek{p})- (m_0/m)
\omegac[0] \, \vek{\hat z}$ is an effective magnetic-field operator that governs spin precession
in the Landau-Rashba model. Similarly, the 2D position operator can be decomposed into
a constant part, which corresponds to the guiding-centre position of a cyclotron orbit,
and a time-dependent oscillatory part. For the sake of brevity, we will use a compact
complex notation~\cite{ahmintro} for 2D position $\vek{r}=(x,y)$ and kinetic-momentum $\vek{\pi}
\equiv \vek{p} + e \vek{A}=(\pi_x, \pi_y)$:
$R = x - iy$, $\Pi = \pi_x - i \pi_y$, and extend this notation also to the in-plane spin
components: $S_\pm = (S_x \pm i S_y)/2$. In the Heisenberg picture, the time
evolution of the complex 2D position is then given by $R(t) = \bar R + \tilde R(t)$, with
\numparts
\begin{eqnarray} \label{eq:barR}
\bar R &=& R - \frac{i \Pi}{m \omegac} \quad , \\ \label{eq:tildeR}
\tilde R(t) &=& \exp \left[-i t \left( \omegac - 
\frac{\vek{\mathcal B}^{\mathrm{JC}} \cdot \vek{S}}{\hbar} \right)\right]
\left[ \cos (\omega_\delta t )  \frac{i \Pi}{m \omegac}
+ \frac{\sin (\omega_\delta t )}{i \omega_\delta} \frac{i \Pi}{m \omegac} \,
\frac{\vek{\mathcal B}^{\mathrm{JC}} \cdot \vek{S}}{\hbar} \right]  .
\nonumber\\
\end{eqnarray}
\endnumparts
Here $\omega_\delta =\sqrt{\left( \vek{\mathcal B}^{\mathrm{JC}} \cdot
\vek{S}/\hbar \right)^2 + \hbar\omegac m \alpha^2/2 }$, and $\omegac=e B/m$ is the cyclotron
frequency of electrons in the semiconductor material.

In principle, the expressions given in \eref{eq:barS}-\eref{eq:tildeR} allow for the
calculation of time-dependent spin and position expectation values for any initial
state. In practice, such a calculation may turn out to be rather cumbersome and difficult
to interpret. Hence, in the following, we will discuss magnetic focusing in the context of
semiclassical approaches applied to the Landau-Rashba model.
Approximate semiclassical approaches are often practical for understanding certain physical
phenomena and also provide a rather general framework to treat quantum systems of interest.
However, the above exact results provide a useful benchmark for their reliability.

\subsection{Precessing-spin semiclassics of the Landau-Rashba model}

In this case, the orbital dynamics is entirely governed by the Landau model. Using the
compact complex notation introduced above, we have
\numparts
\begin{eqnarray}
R(t) &=& \bar R + \frac{i \Pi}{m \omegac}\, e^{-i\omegac t} \quad , \\
\Pi(t) &=& \Pi \, e^{-i\omegac t} \quad .
\end{eqnarray}
\endnumparts
The spin dynamics is determined by the precession equation~\eref{eq:precWeak}, where
$\vek{\mathcal B}$ is given by \eref{eq:laraB}. For the $z$ and in-plane components $s_z$
and $s_\pm = (s_x \pm i s_y)/2$, they read explicitly
\numparts
\begin{eqnarray}\label{eq:nminWeak}
\dot s_- &=& -\frac{\alpha}{2}\, \Pi(t)\, s_z(t) - i \frac{g}{2} \omegac[0] \, 
s_-(t) \quad , \\ \label{eq:nzWeak}
\dot s_z &=& \alpha \left[ s_-(t) \, \Pi^\ast(t) + s_+(t) \, \Pi(t) \right] \quad .
\end{eqnarray}
\endnumparts
[The equation for $s_+(t)$ follows from complex conjugation of \eref{eq:nminWeak}.]
From \eref{eq:nminWeak} and \eref{eq:nzWeak}, it follows that
\numparts
\begin{equation}
{\ddot{\dot s}}_z = - \omega_{\mathrm P}^2 \, \dot s_z \quad ,
\end{equation}
with the spin-precession-related frequency scale
\begin{equation}\label{eq:weakFreq}
\omega_{\mathrm P} = \sqrt{ \left(\alpha \, \pi\right)^2 + 
\omegac^2 \left(1 - \frac{g m}{2 m_0} \right)^2 } \quad .
\end{equation}
\endnumparts
Note that $\pi \equiv |\vek{\pi}| \equiv |\Pi| = \sqrt{2 m E}$.
Integration yields \numparts
\begin{eqnarray}
\dot s_z &=& \mathcal{C}_+ \, e^{i \omega_{\mathrm P} t}
           + \mathcal{C}_- \, e^{-i \omega_{\mathrm P} t}
\quad , \\ \label{eq:weakSol}
s_z(t) &=& s_z
- \mathcal{C}_+ \, \frac{1 - e^{i \omega_{\mathrm P} t}}{i \omega_{\mathrm P}}
+ \mathcal{C}_- \, \frac{1 - e^{-i \omega_{\mathrm P} t}}{i \omega_{\mathrm P}} \, ,
\end{eqnarray}
\endnumparts
with arbitrary constants $\mathcal{C}_\pm$. Inserting this result into \eref{eq:nminWeak} enables
one to find $s_-(t)$. We omit this step here.

It is illustrating to note that $\hbar\omega_{\mathrm P}/2$ emerges as the eigenvalue of
the JC Hamiltonian $\vek{\mathcal B}^{\mathrm{JC}} \cdot \vek{S}$ in the limit where the
kinetic-momentum operators $\pi_x$ and $\pi_y$ are treated as $c$-numbers. This
corresponds to the early semiclassical treatments of the JC model~\cite{foer:jpa:75}.
Thus \eref{eq:weakSol} with \eref{eq:weakFreq} reflects the exact quantum solution for
spin precession in the Landau-Rashba model \eref{eq:tildeS} taken in the appropriate
weak-coupling limit. Our results obtained here generalise those presented in Sec.~5.1.1
of Ref.~\cite{plet:jpa:03} where the weak-coupling limit of the Landau-Rashba model was
previously discussed. The complete disappearance of $\hbar$ from the dynamics described
in this Section is an expected feature of the precessing-spin semiclassical limit. 

\subsection{Adiabatic-spin semiclassics of the Landau-Rashba model}

In the limit where the electron spin is assumed to be either aligned or anti-aligned with the
local field $\vek{\mathcal B}(\vek{r},\vek{p})$ [given by \eref{eq:laraB}], the Landau-Rashba
model specialises to a pair of terms of the form (\ref{eq:adiabat}) that govern the
dynamics of electrons with spin projection $\pm 1/2$. As the Rashba and
Zeeman contributions to $\vek{\mathcal B}(\vek{r},\vek{p})$ are perpendicular to each other, we
have
\begin{equation}
|\vek{\mathcal B}(\vek{r},\vek{p})| = \sqrt{\left(\frac{g}{2}\, \omegac[0] \right)^2 + \alpha^2 \pi^2}
\quad .
\end{equation}
Thus $\pi \equiv |\vek{\pi}| = \sqrt{2 m E} $ is again a constant of the motion. For a fixed value of
conserved energy $E$, it assumes two different values $\pi_\sigma$ for particles distinguished by
spin projection $\sigma/2$, where $\sigma = \pm 1$. The values of $\pi_\sigma$ can be found from
\begin{equation}
\sqrt{\pi_\sigma^2 + \left(\frac{g \omegac[0]}{2 \alpha}\right)^2} = \sqrt{2 m E + \left(m \,\frac{\hbar
\alpha}{2} \right)^2 + \left( \frac{g \omegac[0]}{2 \alpha}\right)^2} - \sigma \, m\, \frac{\hbar\alpha}{2}
\quad .
\end{equation}

The equations of motion resulting from Hamiltonians \eref{eq:adiabat} for the Landau-Rashba
case can be written as
\numparts
\begin{eqnarray} \label{eq:posEOM}
\dot{\vek r} &=& \frac{\vek{\pi}}{m_\sigma} \quad , \\ \label{eq:velEOM}
\dot{\vek{\pi}} &=& -e \, \dot{\vek{r}} \times \vek{B} \quad ,
\end{eqnarray}
which describe cyclotron motion with a spin-dependent effective mass
\begin{equation}
\frac{m_\sigma}{m} =  1 - \sigma\, \frac{m \hbar \alpha/2}{\sqrt{2 m E + (m \hbar \alpha / 2)^2 +
\left[ g \omegac[0] / (2\alpha) \right]^2}} \quad ,
\end{equation}
\endnumparts
and thus spin-dependent frequency $\omegac[\sigma] = e B/m_\sigma$. The
cyclotron radius $r_{\mathrm c} =\pi/(m\omegac)$ is different for the two
spin species because of their different values of $\pi = \pi_\sigma$ for fixed energy $E$. A straightforward calculation yields
\begin{equation} \label{eq:spinCyc}
r_{{\mathrm c}\sigma} = \frac{1}{m \omegac} \sqrt{2 m E
+ 2 \left(m \frac{\hbar\alpha}{2} \right)^2 
- \sigma \, \hbar\alpha\sqrt{2 m E + \left(m \frac{\hbar\alpha}{2}\right)^2
+ \left(\frac{g \omegac[0]}{2\alpha}\right)^2}} .
\end{equation}

A few comments about our results are in order. Firstly, the fact that the above expressions for
spin-dependent cyclotron frequency and radius depend on the parameter $\hbar\alpha/2$
(which has the dimension of velocity)
is a direct consequence of the way the adiabatic-spin semiclassical limit is performed. Secondly,
for $g=0$, our expression for $\omegac[\sigma]$ is exactly the same as that found in
Ref.~\cite{usaj:prb:04} where, ostensibly, the spin-orbit-intertwined semiclassical limit
was discussed. Also, our result for $r_{{\mathrm c}\sigma}$ agrees with the corresponding
expression from Ref.~\cite{usaj:prb:04} to leading order in the large-$E$ limit. Apparently, the
approximate scheme employed by the authors of Ref.~\cite{usaj:prb:04} is essentially equivalent
to the adiabatic-spin semiclassics, even though they recover a finite $z$ component of spin as
given in \Eref{eq:barS} above. Thus a consistent treatment of the Landau-Rashba model
using spin-orbit-intertwined semiclassics seems to be still lacking. Thirdly, the canonical
equations of motion \eref{eq:posEOM} and \eref{eq:velEOM} indicate that the effect of
adiabatic-spin semiclassics on the orbital dynamics is better described as a renormalisation
of the effective mass~\cite{byra:jetplett:84,valin:prb:06} than a renormalisation of the focusing
field~\cite{rokh:prl:04}. Lastly, \eref{eq:velEOM} represents the familiar expression of the Lorentz
force in terms of a particle's velocity {\em without\/} any trace of the previously
claimed~\cite{shen:prl:05} spin-dependent contribution. Such a result is expected from
proper quantum-mechanical derivations~\cite{zawa:07} of the Lorentz-force operator.

\subsection{Analysis of magnetic-focusing experiments}

To leading order, no spin splitting of magnetic-focusing peaks occurs in the precessing-spin
semiclassical limit. For this case, back action of spin dynamics on orbital motion may appear
only in corrections of order $\hbar$ that can be included in principle~\cite{teufel:03}. Instead
of considering this possibility, we focus here on the adiabatic-spin semiclassics which provides
a proper description of magnetic focusing already in leading order.

\begin{figure}[t]
\begin{indented}
\item[] \includegraphics{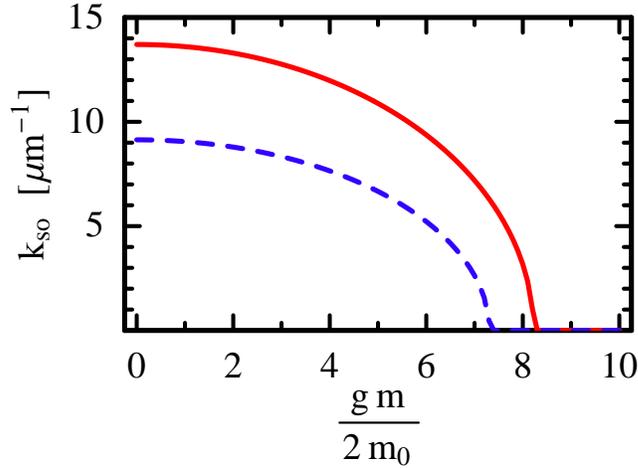}
\caption{\label{fig:expFocGdep}Dependence of extracted $k_{\mathrm{so}}$ on the assumed
value for the reduced $g$-factor $g m/(2 m_0)$. The calculation of the solid red (dashed blue)
curve used focusing data from Ref.~\cite{murph:physe:06} (Ref.~\cite{rokh:prl:04}). Apparently,
a rather weak variation of extracted $k_{\mathrm{so}}$ with $g m/(2 m_0)$ persists to quite
large values of the latter which, in experiment, correspond to comparable magnitudes of Rashba
and Zeeman spin splittings.}
\end{indented}
\end{figure}

It is useful to define effective wave-vector scales $k_{\mathrm F}=\sqrt{2 m E + (m \hbar
\alpha/2)^2}/\hbar$ and $k_{\mathrm{so}}=m\alpha/2$, which are associated with the 2D carrier
sheet density and Rashba spin splitting, respectively. After equating $2 r_{{\mathrm c}\sigma}$
from \eref{eq:spinCyc} with the contact separation $L$ and performing some straightforward
algebra, we find a relation that has to be satisfied by each of the two experimentally observed
focusing fields $B_\pm$:
\begin{equation} \label{eq:magFocRel}
\left( \frac{e L^2 B_\sigma}{4\hbar} - \frac{\hbar}{e B_\sigma} \left[ k_{\mathrm F} - 
k_{\mathrm{so}} \right]^2 \right) \left( \frac{e L^2 B_\sigma}{4 \hbar} - \frac{\hbar}{e B_\sigma}
\left[ k_{\mathrm F} + k_{\mathrm{so}} \right]^2 \right) = \left( \frac{g m}{2 m_0}\right)^2  .
\end{equation}
Thus measurement of the spin-split focusing peaks allows to determine both $k_{\mathrm F}$
and $k_{\mathrm{so}}$ directly, with the effective electron mass entering only the parameter
$g m/(2 m_0)$. For $g=0$, the relation \eref{eq:magFocRel} specialises to $e L B_\sigma = 2
\hbar (k_{\mathrm F} - \sigma k_{\mathrm{so}})$, which was used in Ref.~\cite{rokh:prl:04} to
extract $k_{\mathrm{so}}$ in a GaAs 2D hole system. \Tref{tab:expAnal} summarises values
obtained from existing magnetic-focusing data, taking into account the finite Zeeman splitting.
It turns out that, for the sample parameters realised in these experiments, the dependence of
extracted $k_{\mathrm{so}}$ on $g m/(2 m_0)$ is rather weak over an extended range before
Zeeman splitting becomes suddenly dominant. This surprising feature, which is illustrated in
\Fref{fig:expFocGdep}, explains why it was possible to extract a reasonable value for
$k_{\mathrm{so}}$ in Ref.~\cite{rokh:prl:04} even though, in that experiment, Rashba and
Zeeman spin splittings were of comparable magnitude for states at the Fermi energy.
Note, however, that the range of parameter $g m/(2 m_0)$ over which Rashba splitting can be
reliably extracted will be reduced in samples with smaller contact separation and concomitantly
higher focusing fields.
\Table{\label{tab:expAnal}Parameters associated with and extracted from recent
spin-dependent magnetic-focusing experiments. Besides contact separation $L$ and
focusing fields $B_\pm$, we also provide the value $k_{{\mathrm F}0}=\sqrt{2\pi n_0}$ of
Fermi wave vector as derived from the 2D sheet density $n_0$, which can be compared to
$k_{\mathrm F}$ extracted from the focusing data.}
\br
material & $\frac{g m}{2 m_0}$ & $L$ [nm] & $k_{{\mathrm F}0}$ [nm$^{-1}$] & $B_+$ [T] & $B_-$ [T]
& $k_{\mathrm F}$ [nm$^{-1}$] & $k_{\mathrm{so}}$ [nm$^{-1}$] \\
\mr
p-GaAs$^\dagger$ & 1.4 & 800 & 0.093 &0.17 & 0.20 & 0.11 & 0.009  \\
n-InSb$^\ddagger$ & 0.36 & 600 & 0.14 & 0.30 & 0.36 & 0.15 & 0.014 \\
n-InSb$^\ast$         & 0.36 & 600 & 0.19   & 0.37 & 0.50 & 0.20 & 0.030 \\
\br
\end{tabular}
\item[] $^\dagger$ From Ref.~\cite{rokh:prl:04}. As in this work, we apply the $k$-linear Rashba
			    model for conduction-band electrons to interpret the data. A more detailed study
			    would have to take into account fundamental differences between Rashba spin
			    splitting in 2D electron and hole systems~\cite{rolandbook}.
\item[] $^\ddagger$ From Ref.~\cite{murph:physe:06}.
\item[] $^\ast$ From Ref.~\cite{here:aipproc:07}. The given value of $k_{{\mathrm F}0}$ is derived
		      from the focusing field (0.42 T) expected in the absence of spin splitting (as stated
		      by the authors).
\end{indented}
\end{table}

The values of $k_{\mathrm{so}}$ given in \Tref{tab:expAnal} are on the order of 10\% of the
effective Fermi wave number $k_{\mathrm F}$. Thus the applicability of the adiabatic-spin
semiclassics for typical experimental situations could be questioned. A detailed discussion of
this point would benefit from a fuller understanding of spin-orbit-intertwined semiclassics in the
Landau-Rashba model, which is currently lacking. Incidentally, results from a numerical
simulation~\cite{usaj:prb-rc:04} performed using experimentally realistic parameters provide
strong support for the assumption of adiabatic-spin dynamics.

Our analysis of experimental data focused exclusively on the first magnetic-focusing peak.
This allowed us to neglect scattering at the lithographic barrier between injector and collector
contacts, which is relevant for higher-order focusing peaks. Within a ballistic semiclassical
approach such as ours, spin flips occuring during collision with the sample edge can be taken
into account phenomenologically~\cite{here:apl:05} by including the possibility for particles to
continue on either one of the spin-split cyclotron orbits after each reflection. This model
predicts that the second peak will be unsplit (split into three parts) in the absence (presence)
of boundary spin-flip scattering~\cite{here:aipproc:07}.

\section{Relation to \textit{Zitterbewegung}}

\textit{Zitterbewegung\/} (ZB) was originally introduced by Schr\"odinger as the technical term
for an oscillatory orbital motion performed by free relativistic electrons whose dynamics is
governed by the Dirac equation. See Refs.~\cite{huang:ajp:52,barut:prd:81,barut:prd:85a,tha:92}
for modern descriptions of the effect. ZB has never been directly observed, partly because the
associated period and amplitude ($\lesssim 10^{-21}$~s and $\lesssim 0.004$~\AA, respectively,
for electrons in vacuum) are out of reach for any current experimental equipment. For
ultra-relativistic particles, the ZB amplitude is of the order of the de~Broglie wave
length~\cite{uz:prb:07}, limiting the suitability of scattering experiments to detect the effect.

Analogs of ZB in a nonrelativistic solid-state context have recently attracted great
interest~\cite{zawa:prb:05,schlie:prl:05,schlie:prb:06,jozsef:prb:06,zawa:prb:06,uz:prb:07,zawa:jpcm:07}.
In particular, charge carriers from spin-split bands are expected to perform an oscillatory
motion that is entirely analogous to ZB~\cite{schlie:prl:05,schlie:prb:06,jozsef:prb:06,uz:prb:07}.
The experimentally observed~\cite{crook:prl:05,kato:apl:05b} zero-field spin precession of electrons
and holes turns out to be closely related to ZB~\cite{uz:prb:07}, but no direct ramification of ZB in
coordinate space has been measured. Theoretical studies
suggest~\cite{brus:prb:06,jozsef:prb:06,uz:prb:07} that ZB results in a spatial separation of
carriers with opposite spin that is of the order of the de~Broglie wave length. Here we show
that spin-dependent magnetic focusing is also closely related with ZB. For the sake of
notational simplicity, we neglect Zeeman splitting from now on and consider only the limit of
sufficiently small magnetic fields where Landau-quantisation effects are negligible.

\subsection{Cyclotron motion of charge carriers performing \textit{Zitterbewegung}}

It is a consequence of ZB that the time-dependent position [velocity, spin] operators $\vek{r}(t)$
[$\vek{v}(t)$, $\vek{S}(t)$] can be written as the sum of an average (smoothened over ZB) part
$\vek{\bar r}(t)$ [$\vek{\bar v}(t)$, $\vek{\bar S}(t)$] and an oscillatory part $\vek{\tilde r}(t)$
[$\vek{\tilde v}(t)$, $\vek{\tilde S}(t)$]. Universal expressions, in terms of suitably defined ZB
frequency and amplitude operators $\hat\omega(\vek{p})$ and $\vek{F}$, have been obtained
for all these operators for a range of (multi-band) models~\cite{uz:prb:07}. For example, the
average part of the velocity is given by
\begin{equation}\label{genAvVel}
\vek{\bar v} = \frac{\partial {\mathcal H}}{\partial \vek{p}} - \vek{F}
\end{equation}
for any two-band Hamiltonian ${\mathcal H}$, including the Rashba model. We continue by
discussing this special case only. A finite acceleration of the (ostensibly free!) particles performing
ZB in zero magnetic field is found
\begin{equation} \label{eq:ZBacc}
\dot\vek{v}_{\mathrm{ZB}}(t) = i \, \hat\omega(\vek{p}) \, \vek{F}\, e^{-i \hat\omega(\vek{p}) t}
= \alpha^2 S_z(t) \,\, \vek{p}\times \vek{\hat z} \quad ,
\end{equation}
where the r.h.s\ expression is the specialisation to the Rashba-model case, and we used the
appropriate expression for $\vek{F}$ that, within our current notation, reads
\begin{equation}
\vek{F} = \frac{\partial \left(\vek{\mathcal B}\cdot \vek{S}\right)}{\partial\vek{p}} - \left(\frac{\hbar\alpha}
{2} \right)^2\frac{\vek{p}}{\vek{\mathcal B}\cdot \vek{S}} \quad .
\end{equation}
As the r.h.s\ of \eref{eq:ZBacc} indicates, this acceleration is intimately related to spin precession
in the Rashba model. In particular, the fact that $\bar S_z=0$ in zero magnetic field is directly
associated with the vanishing time average of the ZB-related acceleration \eref{eq:ZBacc} for this
case.

A finite perpendicular magnetic field forces charged particles on cyclotron orbits and, therefore,
leads to additional time dependences of their dynamical variables. To elucidate the interplay
between ZB and classical cyclotron motion more clearly than it emerges, eg, from the exact
solution of the JC model given in Sec.~\ref{JCmodelSec}, we concentrate on the low-field limit
where the time scales associated with these two effects are well-separated.
This regime allows one
to consider quantities that are averaged over the ZB time scale but are still time-dependent
because of the cyclotron motion.

As indicated by \eref{eq:barS}, $\bar S_z$ becomes finite in a perpendicular magnetic field. In
that situation, the average of the acceleration \eref{eq:ZBacc}  over the ZB time scale becomes
finite and yields, in the low-field regime and with Zeeman splitting neglected, 
the Lorentz-force-like term
\begin{equation}\label{eq:ZbLorentz}
\bar{\dot\vek{v}}_{\mathrm{ZB}} (t) = -\frac{e}{m} \, \left(\vek{\bar v}(t) - \frac{\vek{\pi}(t)}{m} \right)
\times \vek{B} \quad .
\end{equation}
To obtain \eref{eq:ZbLorentz}, we used \eref{eq:barS} above, as well as the relation \eref{genAvVel}
specialised to the Landau-Rashba model in the low-field limit (where the non-commutativity of
$\pi_x$ and $\pi_y$ can be neglected),
\begin{equation}
 \label{eq:genAvVel}
\vek{\bar v}  = \left( \frac{1}{m} 
+ \frac{\vek{\mathcal B}\cdot \vek{S}}{\vek{\pi}^2}
\right) \vek{\pi} \quad .
\end{equation}
The total ZB-averaged acceleration experienced by a charged particle subject to
a magnetic field is given, in the low-field limit, by the sum of the ordinary Lorentz-force contribution
and the finite ZB-averaged acceleration \eref{eq:ZbLorentz}. It turns out to have the form
\begin{equation}\label{eq:LorentzZB}
\bar{\dot{\vek{v}}} (t) = -\frac{e}{m} \, \vek{\bar v}(t) \times \vek{B} \quad ,
\end{equation}
ie, is determined by the ZB-averaged velocity $\vek{\bar v} (t)$ derived from \eref{eq:genAvVel}.
Applying the adiabatic-spin semiclassical limit, \eref{eq:genAvVel} specialises
to \eref{eq:posEOM}, and \eref{eq:LorentzZB} is equivalent to \eref{eq:velEOM}. Hence the
spin-dependent cyclotron mass $m_\sigma$ emerges because of the ZB contribution
\eref{eq:ZbLorentz} to the total acceleration \eref{eq:LorentzZB}. Similar to the spatial separation
of spin-polarised partial waves for an unpolarised electron beam injected into a wave
guide~\cite{brus:prb:06}, spin-dependent cyclotron motion is thus a direct consequence of ZB-related
dynamics arising in the presence of spin splitting.

As noted already a long time ago~\cite{barut:prd:81}, ZB-related terms can contribute to expectation
values of observables such as $\expect{\vek{r}^2}$. Similarly, the evaluation of $\expect{\vek{r}^2}$
for eigenstates of the Landau-Rashba model at fixed energy $E$ yields a hint of the existence of
two different cyclotron orbits for electrons from spin-split bands~\cite{usaj:prb:04}.

\subsection{Corrections due to anomalous position operator}

The effective Rashba Hamiltonian describing electrons in the conduction band can be thought of
as arising from a canonical transformation (L\"owdin partitioning~\cite{rolandbook}) that is similar
to the Foldy-Wouthuysen transformation~\cite{foldy:pr:50} needed to arrive at the proper
non-relativistic limit of Dirac-electron theory. That same transformation will change the form of
the {\em physical\/} position operator associated with a point-like particle, which will then be
different from the {\em canonical\/} position operator acting in the reduced Hilbert space of the
conduction band. For Dirac electrons, the two position operators have been proposed to be associated 
with a particle's centre of charge and centre of mass, respectively~\cite{barut:prd:81}. See
Ref.~\cite{noz:jdp:73} for a closely related discussion refering to solid state systems. Transport 
measurements such as the
ones employed by the magnetic-focusing technique can be expected to be sensitive to the
physical (charge) position rather than the canonical (mass) one.

The formal relation between physical position $\vek{r}_{\mathrm p}$ and the canonical one
$\vek{r}$ is
\begin{equation}
\vek{r}_{\mathrm p} = \vek{r} + \frac{2 \Lambda^2}{\hbar^2} \, \vek{\pi} \times \vek{S} \quad ,
\end{equation}
where $\Lambda$ is the (effective) Compton wave length. Typical $\Lambda$-values for carriers
in generic semiconductors can be found in Refs.~\cite{zawa:prb:05,uz:prb:07}. In a finite magnetic
field, the shift between physical and canonical position gives rise to a correction to spin-split
cyclotron radii that is of the order of $|e B| \Lambda^2/\hbar$ and, therefore, usually quite small.
A possible exception could be InSb where $\Lambda \approx 4$~nm.

\section{Conclusions}

We have studied theoretically the cyclotron motion of charge carriers from spin-split bands,
highlighting exact quantum and semiclassical results for electrons in asymmetric 2D
semiconductor heterostructures.  A spin-dependent splitting of cyclotron orbits is found in the
adiabatic-spin semiclassical limit. Relevant parameters of real samples used in recent
magnetic-focusing experiments were extracted, taking full account of Zeeman splitting. Our
analytical formulae should also be useful for analysis and design of future spin-dependent
focusing measurements. We furthermore elucidated the intricate relationship between spin-split
cyclotron orbits and {\em Zitterbewegung\/} of charge carriers in systems with strong spin-orbit
coupling.

Future studies will be aimed at a systematic investigation of similarities and differences
exhibited in the cyclotron motion of particles from generic two-band models such as those
describing relativistic Dirac electrons~\cite{barut:ozjp:82} or holes in a typical semiconductor's
valence band~\cite{luttham}.

\section*{Acknowledgments}

RW and UZ received support from the National Science Foundation under Grant No.\
PHY99-07949 while visiting the Kavli Institute for Theoretical Physics (University of California,
Santa Barbara, USA) where part of this work was performed. The authors thank R~Danneau,
A~R~Hamilton, J~J~Heremans, A~P~Micolich, J~Sinova, and G~Vignale for illuminating
discussions.

\end{document}